\input amstex.tex
\documentstyle{amsppt}
\magnification\magstep1
\NoBlackBoxes

\define\gm{\bold g}
\redefine\a{\alpha}
\redefine\b{\beta}
\redefine\g{\gamma}

\define\tr{\text{tr}}

\hoffset=1cm
\voffset=1cm
\advance\vsize -1cm
\topmatter
\title  Families index theorem in supersymmetric WZW model and twisted
K-theory*   \endtitle

\author Jouko Mickelsson \endauthor 
\affil Department of Mathematics and Statistics,University of
Helsinki; Mathematical Physics,  KTH, Stockholm \endaffil

\endtopmatter

\bf Abstract. \rm The construction of twisted K-theory classes on a compact Lie group is reviewed
using the supersymmetric Wess-Zumino-Witten model on a cylinder. The Quillen superconnection
is introduced for a family of supercharges and the Chern character for the family is given and its relation
to twisted cohomology is discussed.

  \footnote" "{ *Invited talk at  "Problemi Attuali di Fisica Teorica", session on Gerbes and Poisson Geometry,
organized by Paolo Aschieri,  Vietri sul Mare, March 24, 2005 }
\document
\NoRunningHeads
\baselineskip 18pt

\vskip 0.2in

\bf 0.  Introduction \rm 

\vskip 0.2in

Gauge symmetry breaking in quantum field theory is described in  terms of
families index theory.  The Atiyah-Singer index formula gives via the Chern character
cohomology classes in the moduli space of gauge connections and of
Riemann metrics.  In particular,
the 2-form part is interpreted as the curvature of the Dirac determinant line bundle, which
gives an obstruction to gauge covariant quantization in the path integral formalism. The 
obstruction depends only on the K-theory class of the family of operators.

In the Hamiltonian quantization odd forms on the moduli space become relevant, [CMM].
The obstruction to gauge covariant quantization comes from the 3-form part of the
character. The 3-form is known as the Dixmier-Douady class and is also the (only) characteristic
class of a gerbe; this is the higher analogue of the first Chern class (in path integral quantization) 
classifying complex line bundles.  

The next step is to study familes of "operators" which are only projectivly defined; that is, we have  families
of hamiltonians which are defined locally in the moduli space but which refuse to patch to a globally
defined family of operators.  The obstruction is given by the Dixmier-Douady class, an element
of integral third cohomology of the moduli space. On the overlaps of open sets the operators 
are related by a conjugation by a projective unitary transformation. This leads to the definition of
twisted K-theory. 

In the present talk I will review the basic definitions of both ordinary K-theory and twisted K-theory 
in Section 1. In Section 2 the construction of twisted (equivariant) K-theory classes on compact Lie 
groups is outlined using a supersymmetric model in $1+1$ dimensional quantum field theory.
Finally, in  Section 3 the Quillen superconnection formula is applied to the projective family 
of Fredholm operators giving a Chern character alternatively with values in Deligne cohomology
on the base or in global twisted de Rham cocycles, [BCMMS]. 
The use of Quillen superconnection has been proposed in general context of twisted K-theory in [Fr],
but in this talk I will give the details in simple terms using the supersymmetric Wess--Zumino-Witten
model.

\vskip 0.3in

\bf 1. Twist in K-theory by a gerbe class \rm 

\vskip 0.2in

Let $M$ be a compact manifold and $P$ a principal bundle over $M$ with
structure group $PU(H),$ the projective unitary group of a complex Hilbert space $H.$
We shall consider the case when $H$ is infinite dimensional.
The characteristic class of $P$ is represented
by an element $\Omega\in H^3(M,\Bbb Z),$ the Dixmier-Douady class. 

Choose a open cover $\{U_{\alpha}\}$ of $M$ with local transition
functions $g_{\a\b} : U_{\a\b} \to PU(H)$ of the bundle $P.$  

In the case of a good cover we can even choose lifts $\hat
g_{\a\b}:U_{\a\b} \to U(H),$ to the unitary group in the Hilbert space
$H,$ on the overlaps $U_{\a\b}= U_{\a} \cap U_{\b},$ but then we only have
$$\hat g_{\a\b}\hat g_{\b\g} \hat g_ {\g\a} = f_{\a\b\g}\cdot \bold{1} $$
for some $f_{\a\b\g}: U_{\a\b\g}  \to S^1.$

Complex K-theory classes on $M$ may be viewed as homotopy classes of
maps $M\to Fred,$ to the space of Fredholm operators in an
infinite-dimensional complex Hilbert space $H.$ This defines what is
known as $K^0(M).$ The other complex K-theory group is $K^1(M)$ and
this is defined by replacing $Fred$ by $Fred_{*},$ the space of
self-adjoint Fredholm operators with both positive and negative
essential spectrum. 

The twisted K-theory classes are here defined as homotopy classes of
sections of a fiber bundle $\Cal Q$ over $M$ with model fiber equal to
either $Fred$ or $Fred_{*}.$ One sets
$$\Cal Q= P \times_{PU(H)} Fred,$$
and similarly for $Fred_{*},$ where the $PU(H)$ action on $Fred$ is
simply the conjugation by a unitary transformatio $\hat g$
corresponding to $g\in PU(H).$ 

We denote by $K^{*}(M,[\Omega])$ the twisted K theory classes, the
twist given by $P.$  

Using local trivializations a section is given by a family of maps 
$\psi_{\a} : U_{\a} \to Fred$ such that 
$$\psi_{\b}(x) = \hat g_{\a\b}^{-1}(x) \psi_{a}(x)\hat g_{\a\b}(x)$$ 
on the overlaps $U_{\a\b}.$ 

\vskip 0.3in

\bf 2. Supersymmetric construction of $K(M,[\Omega])$ \rm 

\vskip 0.2in

We recall from [M1] the construction the   operator $Q_A$ as a sum of a 'free' supercharge $Q$ and an
interaction term $\hat A$ in (2.7) acting in $H.$ The Hilbert space $H$ is a tensor 
product of a 'fermionic' Fock space $H_f$ and a 'bosonic'  Hilbert 
space $H_b.$ Let $G$ be a connected, simply connected simple compact Lie group of dimension $N$ and $\gm$ its 
Lie algebra. The space $H_b$ carries an irreducible representation 
of the loop algebra $L\gm$ of level $k$ where the highest 
weight representations of level $k$ are classified by a finite set of $G$ 
representations (the basis of Verlinde algebra)  
on the 'vacuum sector'. 

In a Fourier basis the generators of the loop algebra are 
 $T_n^a $ where $n\in\Bbb Z$ and $a=1,..,dim\, G=N.$ 
 
The commutation relations are 
$$[T_n^a,T_m^b]= \lambda_{abc} T_{n+m}^{c} + \frac{k}{4} \delta_{ab}
\delta_{n,-m},\tag2.1$$ 
where the $\lambda_{abc}$'s are  the structure constant of $\gm;$ in the case
when $\gm$ is the Lie algebra of $SU(2)$ the nonzero structure
constants are completely antisymmetric and we use the normalization 
$\lambda_{123}=
\frac{1}{\sqrt{2}},$ corresponding to an orthonormal basis with respect to 
$-1$ times the Killing form. This means that in this basis the Casimir
invariant $C_2= \sum_{a,b,c} \lambda_{abc} \lambda_{acb}$ takes the
value $-N.$ 
 
In an unitary representation of the loop group we have the  hermiticity relations
 
$$(T^a_n)^*= -T^a_{-n}$$
 With this normalization of the basis, for $G=SU(2),$ $k$ is a
nonnegative integer and $2j_0= 0,1,2\dots k$  labels the possible irreducible representations of $SU(2)$ on the
vacuum sector. The case $k=0$ corresponds
to a trivial representation and we shall assume in the following that
$k$ is strictly positive. In general the level $k$ is quantized as
an integer $x$  times twice     the lenght squared of the
longest root with respect to the dual Killing form (this unit is in
 our normalization equal to $1$ in the case $G=SU(2)$); alternatively, we
can write $k= 2x/h^{\wedge},$ where $h^{\wedge}$ is the dual Coxeter
 number of the Lie algebra $\gm.$  

The space  $H_f$ carries an irreducible representations of the
canonical anticommutation relations (CAR), 
$$\psi_n^a \psi_m^b +\psi_m^b\psi_n^a =
2\delta_{ab}\delta_{n,-m},\tag2.2$$ 
and $(\psi^a_n)^* = \psi^a_{-n}.$ The representation is fixed by the 
requirement that there is an irreducible representation of the
Clifford algebra $\{\psi_0^a\}$ in a subspace $H_{f,vac}$ such that 
$\psi_n^a v=0$ for $n<0$ and $v\in H_{f,vac}.$ 

The central extension of the loop algebra at level $2$ is
represented in $H_f$ through the operators 
$$K^a_n = -\frac14 \sum_{b,c; m\in\Bbb Z} \lambda_{abc} \psi_{n-m}^b
\psi_{m}^c, \tag2.3$$ 
which satisfy 
$$[K^a_n, K^b_m]= \lambda_{abc} K^c_{n+m} +\frac12 n \delta_{ab}\delta_{n,-m}.  
\tag2.4$$ 

We set $S^a_n= 1\otimes T^a_n + K^a_n \otimes 1.$ This gives a representation of the loop
algebra; in the case $G=SU(2)$ the  level is $k+2$ in the tensor
product $H=H_f\otimes H_b.$ In the parametrization of the level by the
integer $x$ this means that we have a level shift $x\mapsto
x'=x+h^{\wedge}.$

Next we define 
$$Q= i\sum_{a,n} \left( \psi^a_n T^a_{-n} +  \frac{1}{3} \psi^a_n K^a_{-n}\right).\tag2.5$$ 
This operator satisfies $Q^2 =h,$ where $h$ is the hamiltonian of the 
supersymmetric Wess-Zumino-Witten model, 
$$h= - \sum_{a,n} : T^a_n T^a_{-n} : + 
\frac{k+2}{8} \sum_{a,n} :n \psi^a_n\psi^a_{-n}: +\frac{N}{24},\tag2.6$$
where the normal ordering $::$ means that the operators with negative
Fourier index are placed to the right of the operators with positive
index,  $:\psi^a_{-n} \psi^b_{n}:\,\, = -\psi^b_{n} \psi^a_{-n}$ if $n>0$ 
and $:AB: = AB$ otherwise. In the case of the bosonic currents $T^a_n$ 
the sign is $+$ on the right-hand-side of the equation.

Finally, $Q_A$ is defined as 
$$Q_A = Q + i\tilde k \sum_{a,n} \psi^a_n A^a_{-n}\tag 2.7$$ 
where the $A^a_n$'s are the Fourier components of the $\gm$-valued 
function $A$ in the basis $T^a_n$ and $\tilde k = \frac{k+2}{4}.$

The basic property of the family of self-adjoint Fredholm operators
$Q_A$ is that it is equivariant with respect to the action of the 
central extension of the loop group $LG.$ Any element $f\in LG$ is 
represented by a unitary operator $S(f)$ in $H$ but the phase of $S(f)$ 
is not uniquely determined. The equivariantness property is
$$S(f^{-1}) Q_A S(f) = Q_{A^f}\tag2.8$$ 
with $A^f = f^{-1} A f +f^{-1} df.$ For the Lie algebra we have the relations     

$$[S^a_n, Q_A] = i{\tilde k} ( n\psi^a_n + \sum_{b,c; m} \lambda_{abc}\psi^b_m A^c_{n-m})  
\tag2.9$$

The group $LG$ can be viewed as a subgroup of the group $PU(H)$
through the projective representation $S.$ The space $\Cal A$ of smooth 
vector potentials on the circle is the total space for a principal 
bundle with fiber $\Omega G \subset LG,$  the group of based loops at $1.$
Since now $\Omega G \subset PU(H),$ 
$\Cal A$ may be viewed as a reduction of a $PU(H)$ principal bundle 
over $G.$ The $\Omega G$ action by conjugation on the Fredholm
operators in $H$ defines an associated fiber bundle $\Cal Q$ over 
$G$ and the family of operators $Q_A$ defines a section of this  
bundle. Thus $\{Q_A\}$ is a twisted K-theory class over $G$ where
the twist is determined by the level $k+2$ projective
representation of $LG.$  

Actually, there is additional  gauge symmetry due to constant global gauge
transformations. For this reason the construction above leads to
elements in the G-equivariant twisted K-theory $K^*_G(G, [\Omega]),$
where the G-action on $G$ is the conjugation by group elements.
It happens that in the case of $SU(2)$ the construction gives all generators for both equivariant and
nonequivariant twisted K-theories, but not  for other
compact Lie groups. 
 
\vskip 0.3in

\bf 3. Quillen superconnection \rm \newline

Let $Q_A$ be the supercharge associated to the vector potential $A$ on
the circle, with values in the Lie algebra $\gm.$ Recall that this
transforms as
$$ \hat g^{-1} Q_A \hat g= Q_{A^g} $$
with respect to $g\in LG.$ Consider the trivial Hilbert bundle over
$\Cal A$ with fiber $H,$ the operators $Q_A$ acting in the fibers. 
Define a covariant differentiation $\nabla$ acting on the sections of 
the bundle, $\nabla= \delta +\hat\omega$ where $\delta$ is the exterior
differentiation on $\Cal A$ and $\hat\omega$ is a connection 1-form
defined as follows. First, any vector potential on the circle can be 
uniquely written as $A=f^{-1}df$ for some smooth function $f:[0,2\pi] \to 
G$ such that $f(0)=1.$ A tangent vector at $f$ is then represented by
a function $v:[0,2\pi] \to \gm$ such that $v(0)=0$ with periodic
derivatives at the end points, $v=f^{-1}\delta f.$  We set
$$\omega_f(v)= v - \alpha(x)  f(x)^{-1} (\delta f(2\pi) f(2\pi)^{-1})
f(x)\tag3.1$$ 
where $\alpha$ is a fixed smooth real valued function on $[0,2\pi]$ such
that $\alpha(0)=0, \alpha(2\pi)=1$ and all derivatives equal to zero 
at the end points. The meaning of the second term in (3.1) is that it
makes the whole expression periodic so that $\omega$ takes values in 
$L\gm.$ Then $\hat{\omega}_f(v)$ is defined by the projective
representation $S$ of $L\gm$ in $H.$ 

The gauge transformation $A\mapsto A^g$ corresponds to the right translation 
$r_g(f)= fg,$ which sends $\omega_f(v)$ to $g^{-1}\omega_f(v) g.$ 
However, for the quantized operator $\hat \omega$ we get an additional
term. This is because of the central extension $\widehat{LG}$ which acts 
on $\hat\omega$ through the adjoint representation. One has
$$\hat g^{-1} \hat\omega \hat g= \widehat{g^{-1}\omega g} +
  \gamma(\omega,g)$$ 
with
$$\gamma(\omega,g)= \frac{k+2}{8\pi} \int_{S^1}  <\omega, dg
g^{-1}>_K.$$
The bracket $<\cdot,\cdot>_K$ is the Killing form on $\gm.$  
But one checks that the modified 1-form
$$\hat\omega_c=\hat \omega - \frac{k+2}{8\pi} \int_{S^1} <\omega, f^{-1}df>_K$$ 
transforms in a linear manner, 
$$\hat g^{-1} \hat\omega_c \hat g = \widehat{ r_g \omega_c}.\tag3.2$$ 

Here $r_g$ denotes the right action of $\Omega G$ on $\Cal A$ and the
induced right action on connection forms.
We would like to construct characteristic classes on the quotient 
space $\Cal A/\Omega G$ from classes on $\Cal A$ using the
equivariantness property (3.2). 
First, we can construct a Quillen superconnecton [Qu] as the mixed form
$$D=\sqrt{t} Q_A + \delta + \hat\omega_c  - \frac{1}{4\sqrt{t}}  <\psi,f>, \tag3.3$$
where $f$ is the Lie algebra valued curvature form computed from
the connection $\omega$ and $<\psi,f>= \sum \psi^a_n f^a_{-n}.$ 
Formally, this expression is the same as the Bismut-Freed superconnection
for families of Dirac operators, [Bi]. Here $t$ is a free positive real scaling parameter.
This is introduced since in the case of Bismut-Freed superconnection one obtains
the Atiyah-Singer families index forms in the limit $t\to 0$ from the
formula (3.4) or (3.5) below.

We define 
a family of closed differential forms on $\Cal A$ from
$$\Theta = \tr_s \, e^{- ( \sqrt{t} Q_A + \delta + \hat\omega_c -\frac{1}{4\sqrt{t}} <\psi,f>)^2}, \text{ dim$\,G$
 even} \tag3.4$$
In the case when dim$\,G$ is even the supertrace is defined as
$\tr_s(\cdot) = \tr\, \Gamma(\cdot).$ Here $\Gamma$ is the grading
 operator with eigenvalues $\pm 1.$ It is defined uniquely up to a phase $\pm
 1$ by the requirement that it anticommutes with each $\psi^a_n$ and
 commutes with the algebra $T^a_n.$ 

To get integral forms the
 $n$-form part of $\Theta$ should be multiplied by $(1/2\pi i)^{n/2}.$ In the odd case the above
 formula has to be modified: 
$$\Theta = \tr^{\sigma} \, e^{-(\sigma \sqrt{t} Q_A+ \delta + \hat\omega_c -\frac{1}{4\sqrt{t}}  <\psi,f>)^2},\tag3.5$$   
where $\sigma$ is an odd element, $\sigma^2=1,$  anticommuting with odd differential
forms and commuting with $Q_A,$ and the trace $\tr^{\sigma}$ extracts
the operator trace of the coefficients of the linear term in $\sigma.$  In this case the $n$-form part
should be multiplied by $\sqrt{2i} (1/2\pi i)^{n/2}.$ 

The problem with the expressions (3.4) and (3.5) is that they cannot
pushed down to the base $G=\Cal A/\Omega G.$ The obstruction comes from
the transformation property
$$\gather  \sqrt{t} Q_{A^g} + \delta + \widehat{r_g \omega_c }  -\frac{1}{4\sqrt{t}}  <\psi,r_g f>   \\
 = \hat g^{-1} (\sqrt{t} Q_A +\delta +
\hat\omega_c -\frac{1}{4\sqrt{t}}  <\psi,f>)\hat g  + \hat g^* \theta, \tag3.6 \endgather $$ 
where $\theta$ is the connection 1-form on $\widehat{LG}$
corresponding to the curvature form $c$ on $LG,$ defined by the
central extension. Here $\hat g$ is a local $\widehat{LG}$ valued
function on the base $G,$ implementing a change of local section 
$G \to \Cal A.$ This additional term is the difference 
$$\hat g^* \theta= \widehat{g^{-1}\delta g} - \hat g^{-1}\delta\hat
g,$$ 
where the first term on the right comes from the tranformation of the 
connection form $\hat\omega_c$ with respect to a local gauge
transformation $g.$ 
Taking the square of the transformation rule (3.6) we get 
$$(r_g D)^2 = \hat g^{-1} D^2 \hat g + g^* c.\tag3.7$$
The last term on the right comes as 
$$\delta \hat g^*\theta + (\hat g^* \theta)^2 = \hat g^* \delta \theta 
= \hat g^* c = g^* c,$$
where in the last step we have used the fact that the curvature of a
circle bundle is a globally defined 2-form $c$ on the base, and thus
does not depend on the choice of the lift $\hat g$ to $\widehat{LG}.$ 

\proclaim{Theorem} Let $U_{\a}$ and $U_{\b}$ be two open sets in $G$ 
with local sections $\psi_{\a},\psi_{\b}$ to the total space $\Cal A$ of the 
$\Omega G$ principal bundle $\Cal A\to G.$ Let $g_{\a\b}: U_{\a}\cap
U_{\b} \to \Omega G$ be the local gauge transformation tranforming
$\psi_{\a}$ to $\psi_{\b}.$ Then the pull-back forms $\Theta_{\a}$ 
and $\Theta_{\b}$ are related on $U_{\a}\cap U_{\b}$ as 
$$\Theta_{\b} =\psi^*_{\b}\Theta = e^{-g^*_{\a\b}c} \Theta_{\a}.$$ 
\endproclaim
\demo{Proof} Since the curvature is closed, $\delta c=0,$ the term
$g^* c$ on the right in (3.7) commutes with the rest and therefore can 
be taken out as a factor $\exp(-g^* c)$ in the exponential of the
square of the transformed superconnection.  \enddemo  

\bf Remark \rm It is an immediate consequence of the Theorem that the 
1-form part $\Theta[1]$ of $\Theta$ is a globally defined form on the 
base $G.$ 
We can view this as the generalization of the differential
of the families $\eta$ invariant, governing the spectral flow along 
closed loops in the parameter space; in fact, in the classical case 
of Bismut-Freed superconnection for families of Dirac operators this 
is exactly what one gets from the Quillen superconnection formula. 
We can write $\pi^{-1/2} \Theta[1]= h^{-1}dh/2\pi i$ with $\log h= 2\pi i \eta.$ Note that
$\eta$ is only continuous modulo integers.  Thinking of $\eta$ as the spectral asymmetry,
we normalize it by setting $\eta(A)=0$ for the vector potential $A=0,$ or on the base,
for the trivial holonomy $g=1.$

In the odd case we can relate the calculation of $\Theta[3]$ to the 
computation of the Deligne class in twisted K-theory, [Mi2]. 

On the overlap $U_{\a\b}$ we have from the Theorem:
$$\Theta_{\b}[3] =\Theta_{\a}[3] -g^*_{\a\b}c \wedge \Theta_{\a}[1].\tag2.8$$ 
 This gives
$$ \Theta_{\a}[3] -\Theta_{\b}[3]= d(\hat g^*_{\a\b}\theta \wedge
\Theta[1])
\equiv d\omega_{\a\b}^2.\tag3.9$$
Using $\hat g_{\a\b}\hat g_{\b\g} \hat g_{\g\a} = f_{\a\b\g}$ we get  
$$\omega_{\a\b}^2 -\omega_{\a\g}^2 + \omega^2_{\b\g}= (f^{-1}_{\a\b\g} 
df_{\a\b\g})\wedge \Theta[1].\tag3.10$$ 

Choose a function $h:G\to S^1$ as in the Remark,  $\pi^{-1/2} \Theta[1]= h^{-1}dh/2\pi i.$ 

Next the  \v{C}ech coboundary of the cochain $\{\omega^2_{\a\b}\}$ in (3.10) 
can be written as
$$d\omega^1_{\a\b\g} = d( \log(f_{\a\b\g}) h^{-1} dh). \tag3.11$$ 
Defining $a_{\a\b\g\delta}= \log(f_{\b\g\delta}) -   \log(f_{\a\g\delta})
+ \log(f_{\a\b\delta}) -\log(f_{\a\b\g})$ we can write 
$$(\partial \omega^1)_{\a\b\g\delta}  =
h^{-a_{\a\b\g\delta}}dh^{a_{\a\b\gamma\delta}}\tag3.12$$
where $\partial$ denotes the \v{C}ech coboundary operator.
Thus the collection \newline
$\{\Theta_{\a}[3], \omega^2_{\a\b},
\omega^1_{\a\b\g}, h^{a_{\a\b\g\delta}}\}$ defines a Deligne cocycle
on the manifold $G$ with respect to the given open covering
$\{U_{\a}\}.$ 

The system of closed local forms obtained from the Chern character formula (3.5) can be modified
to a system of  \it global forms  \rm by multiplication 
$$ \Theta' = e^{- \theta_{\a}}  \wedge \Theta, \tag3.13$$
where $\theta_{\a}$ is the 2-form potential, $d\theta_{\a} =\Omega$ on $U_{\a}.$  
One checks easily that now
$$(d+\Omega) \Theta' =0. \tag3.14$$

Although the operator $d +\Omega$ squares to zero and can thus be used 
to define a cohomology theory, [BCMMS],  the Chern character should not be viewed to give a map
to this twisted cohomology theory. In fact, the twisted cohomology over complex numbers vanishes for simple
compact Lie groups. For this reason, in order to hope to get nontrivial information from
the Chern character, one should look for a refinement of the twisted cohomology.
In fact, there is another integral version of twisted cohomology proposed  in [At]. In that version one
studies the ordinary integral cohomology modulo the ideal generated by the Dixmier-Douady class
$\Omega.$ At least in the case of $SU(2)$ it is an experimental fact that the twisted (nonequivariant) 
K-theory as an abelian
group is isomorphic to the twisted cohomology in this latter sense.
There is a similar result for
other compact Lie groups, but the 3-cohomology class used to define the twisting in cohomology 
is in general not the gerbe class; both are integral multiples of a basic 3-form, but the coefficients
differ, except for the case of $SU(2),$ [Do].   

One can explicitly see why the integral cohomology mod $\Omega$ is relevant for twisted K-theory
by the following construction in the odd case.  First we replace the space $Fred_*$ by the homotopy equivalent
space $\Cal{U}_1$ consisting of $1+$ trace-class unitaries in $H.$ An ordinary K-theory class on $X$ is a homotopy
class of maps $f:X \to \Cal{U}_1.$ In this representation the Chern character  defines a sequence of cohomology classes on $X$ by pulling back the generators
$\tr\, (g^{-1}dg)^{2n+1}$ of the cohomology of $\Cal{U}_1.$  In the twisted case we have only maps on open sets $U_{\a}$ which are related 
by $f_{\b} = g_{\a\b}^{-1} f_{\a} g_{\a\b}$ on overlaps.

In the case of $G=SU(2)=S^3$ we need only two open sets $U_{\pm},$ the slightly extended upper and lower hemispheres, and a map 
$g_{-+}$ on the overlap to the group $PU(H),$ of degree $k.$ 
If now $f_-=1$ identically and the support of $f_+$ is concentrated at the North pole, then the pair $f_{\pm}$ is related by the
conjugation by $g=g_{-+}$ at the equator and at the same time it defines an ordinary K-theory class since the functions
patch to a globally defined function on $S^3.$ Let us assume that the winding number of $f: S^3 \to \Cal{U}_1$ is $k.$ 
Next form a continuous path $f_{\pm}(t)$ of representatives of twisted K-theory classes starting from $f(1)=f$ and ending at 
the trivial class represented by the constant function $f(0)=1.$ Let $\rho$ be a smooth function on $S^3$ which is equal to
$1$ on the overlap $U_{-+}$ and zero in small open neighborhoods $V_{\pm}$  of the poles. We can also extend the domain of definition of $P(x)$ 
to a larger set  $U_+\setminus V_+.$ 
For $0\leq t \leq 1$ define 
$$ f_-(t)= e^{2\pi it \rho(x)P_0}
 \text{ and } f_+(t) = e^{  2\pi  it \rho(x)  P(x) }$$
where $P(x)= g^{-1}(x) P_0 g(x),$ with $P_0$ is a fixed rank one projection, $x\in U_{-+}.$ 
These are smooth functions on $U_{\pm}$ respectively and are related by the conjugation by $g$ on the overlap.
But for $t=0$ both are equal to the identity $1\in \Cal{U}_1.$ On the other hand, at $t=1$ the integral 
$$\frac{1}{24 \pi^2} \int_{S^3} \tr\, (f^{-1} df)^3$$
is easily computed to give the value $k.$  This paradox is explained by the fact that for the intermediate values
$0 < t < 1	$ the functions $f_{\pm}$ do not patch up to a global function on $S^3.$  Thus we have a homotopy joining
a pair of (trivial) twisted K-theory classes corresponding to the pair of third cohomology classes $0, \Omega$ 
computed from the Chern character. This confirms the claim, at least in the case of $X=S^3,$ that the values 
of the Chern character should be projected to the quotient $H^*(X,\Bbb Z)/\Omega.$ 

\vskip 0.3in

\bf References \rm 

\vskip 0.2in
[At] M. Atiyah: K theory past and present.  math.KT/0012213.
Sitzungsberichte der Berliner  Mathematischen Gesellschaft,  411--417, Berliner Math. Gesellschaft, Berlin, 2001.

[AS] M. Atiyah and G. Segal:  Twisted K-theoory. math.KT/0407054
 
[Bi] J.-M. Bismut: Localization formulas, superconnections, and the index
 theorem for families.   Comm. Math. Phys.  \bf 103, \rm  no. 1, 127--166 (1986) 

[BCMMS] P. Bouwknegt, A. Carey,  V. Mathai,  M. K. Murray, and D. Stevenson: Twisted K-theory and K-theory of
bundle gerbes.  0106194. Commun. Math. Phys.  \bf 228, \rm  17-49 (2002)

[CMM] A. Carey, J. Mickelsson, and M.K. Murray: Index theory, gerbes,
and hamiltonian quantization hep-th/9511151
Commun.Math.Phys.\bf 183, \rm 707-722 (1997)

[Do] Christopher L. Douglas:  On the twisted K-homology of simple Lie groups.  math.AT/0402082

[Fr]  D. Freed: Twisted K-theory and loop groups. math.AT/0206237. Publ. in the proceedings of
ICM2002, Beijing.

[FHT] D. Freed, M. Hopkins, and C. Teleman: Twisted  equivariant K-theory with complex coefficients.
math.AT/0206257

[Mi1]  J. Mickelsson: Gerbes, (twisted) K theory, and the supersymmetric WZW model. hep-th/0206139.
Publ. in  \it Infinite Dimensional Groups and Manifolds,   \rm  ed. by T. Wurzbacher. IRMA Lectures in
Mathematics and Theoretical Physics \bf 5, \rm Walter de Gruyter \& , Berlin (2004)

[Mi2]  J. Mickelsson: Twisted K theory invariants.  math.AT/0401130. Lett. in Math. Phys. \bf 71, \rm  109-121 (2005)

[Qu] D. Quillen: Superconnections and the Chern character.  Topology
\bf 24, \rm  no. 1, 89--95 (1985)

\enddocument